# An immuno-Infrared Sensor detects preclinical Alzheimer's and Parkinson's Disease by misfolding

Grischa Gerwert[1,2], Marvin Mann[1,2], Lennart Langenhoff[1,2], Nathalie Woitzik[1,2], Diana Hubert[1,2], Deniz Duman[1,2], Adrian Höveler[1,2], Sandy Budde[1,2], Jonas Simon[1,2], Léon Beyer[1,2], Martin Schuler[1,2], Sandrina Weber[3], Brit Mollenhauer[3,4], Carsten Kötting[1,2], Jörn Güldenhaupt[1,2] and Klaus Gerwert[1,2]*

[1]Center for Protein Diagnostics (PRODI), Ruhr-University Bochum, Bochum, Germany; [2]Department of Biophysics, Ruhr-University Bochum, Bochum, Germany; [3]Department of Neurology, University Medical Center Goettingen, Goettingen, Germany; [4]Paracelsus-Elena-Klinik, Kassel, Germany; *klaus.gerwert@rub.de

*Abstract*

*The immuno-infrared sensor detects target proteins in solution. Exemplary, the initial misfolding of amyloid beta (A$\beta$) peptides in blood is measured, enabling early risk prediction of Alzheimer's disease in the preclinical stage. Antibodies concentrate the target protein on the functionalized attenuated total reflection crystal surface. A quantum cascade laser is used to measure the amide I band, which indicates the secondary structure distribution of A$\beta$ in blood as a biomarker.*

Very early diagnosis of Alzheimer´s disease (AD) is crucial for the successful application of therapeutic drugs, which have recently received FDA-approval.[1] AD starts in a symptom-free stage characterized by the misfolding of amyloid beta (Aβ) peptides from alpha helical/random coil secondary-structures to β-sheet. This conformational change leads first to oligomer formation, then to larger fibrils, and ultimately to macroscopic amyloid plaques in the brain (Figure 1a). Typical clinical symptoms only emerge at late stages when brain damage is largely irreversible and therapeutic response is therefore limited. Hence, the development of early-stage blood-based biomarkers remains a significant unmet medical need.

We have demonstrated that the initial Aβ misfolding in blood can indicate quantitative AD risk during the preclinical, symptom-free stage.[2–4] The misfolding of Aβ is measurable from blood or cerebral fluid (CSF) with a novel optical biosensor, the immuno-infrared sensor (iRS).[5] In contrast, currently established biomarkers measure the reduction of Aβ concentration in body fluids - but only during the symptomatic clinical stages.[6] Due to the accumulation of Aβ in amyloid plaques their concentration decreases in external body fluids. In addition, Aβ aggregation eventually triggers the phosphorylation of p-Tau217, another blood-based biomarker that also appears late, when the effectiveness of drugs is already limited. Thus, the unmet medical need of early diagnostics is now addressed by the iRS assay, which analyzes initial misfolding instead of later concentration decrease.

**The iRS Concept**

The new infrared spectroscopy-based iRS platform enables early quantitation of the Aβ-misfolding in blood and CSF by overcoming key challenges inherent in traditional infrared spectroscopy. The use of quantum cascade lasers (QCL)[7,8] with intense, small diameter IR beams enables in contrast to the weak, broadband light sources of conventional Fourier-transform infrared (FTIR) instruments, parallel measurements and thereby scalability. This is a prerequisite for clinical applications.

A fundamental challenge in mid-IR analysis is detecting the secondary structure sensitive amide I absorbance band of very low concentration biomarkers amidst a strong background absorbance from water. An amide I band at 1652 cm⁻¹ corresponds to alpha-helical or random coil structures, while a down shift indicates misfolding to β-sheet structures (Figure 1b).[9] Overlapping water absorbance— specifically the OH bending vibration near 1640 cm⁻¹—is several orders of magnitude larger, complicating detection. This is mitigated by attenuated total reflection (ATR) spectroscopy, where the evanescent IR wave penetrates only a ~500 nm thin surface layer of the fluid sample flown over the ATR crystal (Figure 1c) reducing largely the water background absorbance.[10] Difference spectroscopy then precisely subtracts the water background, isolating the biomarker signal (Figure 2). The difference spectrum before (1) and after capture (2) of the biomarker from a body fluid provides the absorbance maximum of the biomarker as a read-out. The background absorbance of all surface-bound molecules including water cancel out in the difference. The maximum of the amide-I band indicates at 1652 cm$^{-1}$ mostly monomers for healthy individuals and in this example at 1624 cm$^{-1}$ mostly fibrils for diseased. During progression of the disease the band downshifts by misfolding to oligomers and fibrils in the symptom free stage, indicating a high risk for a later clinical diagnosis.

The approach of infrared difference spectroscopy has already proven successful in time-resolved FTIR studies investigating protein reaction mechanisms at atomic resolution (Figure 3). The expertise obtained in time-resolved infrared spectroscopy of proteins was key in overcoming the challenges in the development of the iRS. Subsequently this approach was the proven successfully to work in an ATR experiment to reveal minor signals in a large water background.[11]

**Covalent surface modification**

To ensure that only the biomarker is captured and concentrated on the ATR surface, it is functionalized with target-specific capture molecules—here, antibodies (Figure 1c). These antibodies selectively concentrate the biomarker within the thin IR-probed layer, conferring the sensor with the specificity and sensitivity of the capture antibody. However, unspecific binding of other fluid components would also generate absorbance, indistinguishable from specific signals. To overcome this, we developed a biorthogonal surface chemistry that blocks non-specific adsorption and covalently binds antibodies. This surface functionalization is a key innovation of the iRS assay (Figure 4). All biochemical reaction steps during functionalization are monitored in real time by IR spectroscopy to ensure reproducibility and consistent quality control. Figure 4a shows the chemical steps for the preparation of the iRS surface and Figure 4b shows the complete surface.[12–14] First, the silicon ATR crystal is cleaned, polished and the surface is oxidized in a plasma cleaner to obtain a silicon oxide layer. In the following attachment step, triethoxysilane-based linker molecules **1** are attached to this layer.[15] The ethoxy groups are hydrolyzed under defined conditions to form reactive silanols that can condense on the surface of the oxidized silicone. As a second reactive group, linker molecule **1** has an azide group. The latter can be used for strain-promoted click chemistry[16,17] with the second linker molecule **2**, which has an aza-dibenzocyclooctyne (DBCO)[18] moiety. The second reactive group of **2** is N-hydoxysuccinimide (NHS). This group is reactive to primary amines, as found in peptides within the lysine side chains and the N-terminus.[19] The reaction with partially lysed casein, which is commonly used for noncovalent blocking, results here in a covalent blocking layer. Next, this blocking layer must be functionalized. Again, a linker with NHS (**3**) is used to covalently attach the linker to further amine groups of the blocking layer. The second functional group of linker **3** is again an azide. This azide can be used for click-chemistry with DBCO. To immobilize antibodies, they must be modified with DBCO, e.g., by attaching another NHS-DBCO linker (**4**) to an amine group or by using maleimido-DBCO for attachment to a cysteine residue.[20] To achieve good solubility and flexibility, and to avoid hydrophobic regions prone to unspecific binding, the two reactive groups of each linker molecule should be separated by several PEG groups.. The resulting surface (Figure S2B) is stable because all the molecules are covalently bound, and the complete blocking layer effectively prevents unspecific binding.

**Body-fluid measurement with a QCL-based iRS instrument**

The performance of the QCL based iRS is shown in Figures 5 & 6. For surface functionalization and measurements of body-fluids, we developed a fluidic system with an advanced flow chamber as exchangeable disposable positioned on top of the ATR crystal surface to allow precise, reproducible fluid handling. Four channels on two ATR crystals can be measured in parallel. During measurement, the disposable is first filled with buffer to record the background absorbance of the functionalized surface and water. Then, the body fluid sample flows over the surface, selectively capturing the biomarker. Subtraction of the dominant background reveals the amide I and amide II bands of the biomarker (Figure 5). The amide I frequency provides a readout of the secondary structure distribution, while the amide II band relates to biomarker concentration. Figure 5a shows the kinetics of the amide II band during antibody immobilization. The amide II band indicates the amount of antibody bound to the surface within few minutes. After antibody binding a buffer washes off non-covalently absorbed antibody from the surface. Finally, the absorbance of the amide II band is very similar in all four channels indicating a very similar amount of antibody covalently bound on the surface. Amide I and II bands of the catcher antibody in four different channels on two different crystals are shown in Figure 5b. In the next step, the body fluids are flushed over the surface. In the resulting difference spectra (Fig 5 c,d) amide I and II bands of a PD (Parkinson Disease) CSF sample and a DC (Disease Control) CSF sample reflect the binding of alpha-synuclein on the surface. The amide I band of the DC sample has an absorbance maximum at 1643.1 $cm^{-1}$ while the maximum of the PD sample amide I band is at 1638.2 $cm^{-1}$ as expected for healthy and PD patients, respectively.

**Comparison of FTIR and QCL based iRS instruments**

Figure 6 compares the signal to noise ratio of a FTIR based instrument with a globar as the light source and a QCL-based iRS instrument. For this purpose, a difference spectrum of water was measured with both setups under the same conditions. In the amide I region around 1650 $cm^{-1}$, the noise is mainly influenced by the large background absorption of the bending vibration of water. The root mean square (RMS) of the noise in this region is $2.5 \times 10^{-5}$ for the QCL and $3.1 \times 10^{-5}$ for the FTIR system. The noise in the OH bending region was reduced by dynamic calibration of the QCL system. In this mode, the laser intensity is increased in the spectral region of the larger background absorbance of $H_2O$. This provides an improved S/N in the crucial amide I spectral region of the QCL based iRS spectra. The iRS instrument is shown in detail in Figure 7.

**Current applications of the iRS**

In a large population-based cohort, blood plasma samples from a subset of 10,000 participants were analyzed. The iRS assay predicted AD risk up to 17 years before clinical onset during the symptom-free stage, achieving an area under the curve (AUC) of up to 0.82.[2,5] In a cohort with subjective cognitive declined subjects, it predicted conversion to clinical AD six years in advance with an AUC of 0.94 in blood plasma.[3]

More recently, the iRS platform is applied to Parkinson's disease[21] by detecting misfolded α-synuclein in cerebrospinal fluid. In a discovery and independent validation study, sensitivity and specificity were 94% and 97%, respectively. Preliminary unpublished data extend these findings to blood samples. A feasibility study also classified ALS by detecting misfolded TDP-43 in CSF.[22]

Currently, the iRS platform is applied in clinical studies to monitor therapeutic drug effects. Developing a preventive blood screening test for AD and PD accessible for the general population represents the next major milestone. Beyond proteopathies, the iRS has broad potential for chemical and biochemical process and quality control applications.

**Author Contributions**

Conceptualization: C.K., J.G., K.G.

Methodology: G.G., M.M., L.L., N.W., D.H., D.D., A.H., S.B., J.S., L.B., M.S., C.K., J.G., K.G.

Software: J.G.

Investigation&Data Curation: G.G., M.M., L.L., N.W., D.H., D.D., A.H., S.B., J.S., L.B., M.S.

Resources: G.G., J.G., B.M., S.W.

Supervision: C.K., J.G., K.G.

Funding Acquisition: K.G.

G.G., C.K. and K.G. wrote the initial version of the manuscript.

All authors reviewed the manuscript.

**Funding**

The research presented was funded by the Center for Protein Diagnostics (PRODI), Ministry of Culture and Science of North-Rhine Westphalia.

**Competing Interest**

K.G. is the founder and CEO of betaSENSE GmbH. The remaining authors declare no competing interests.

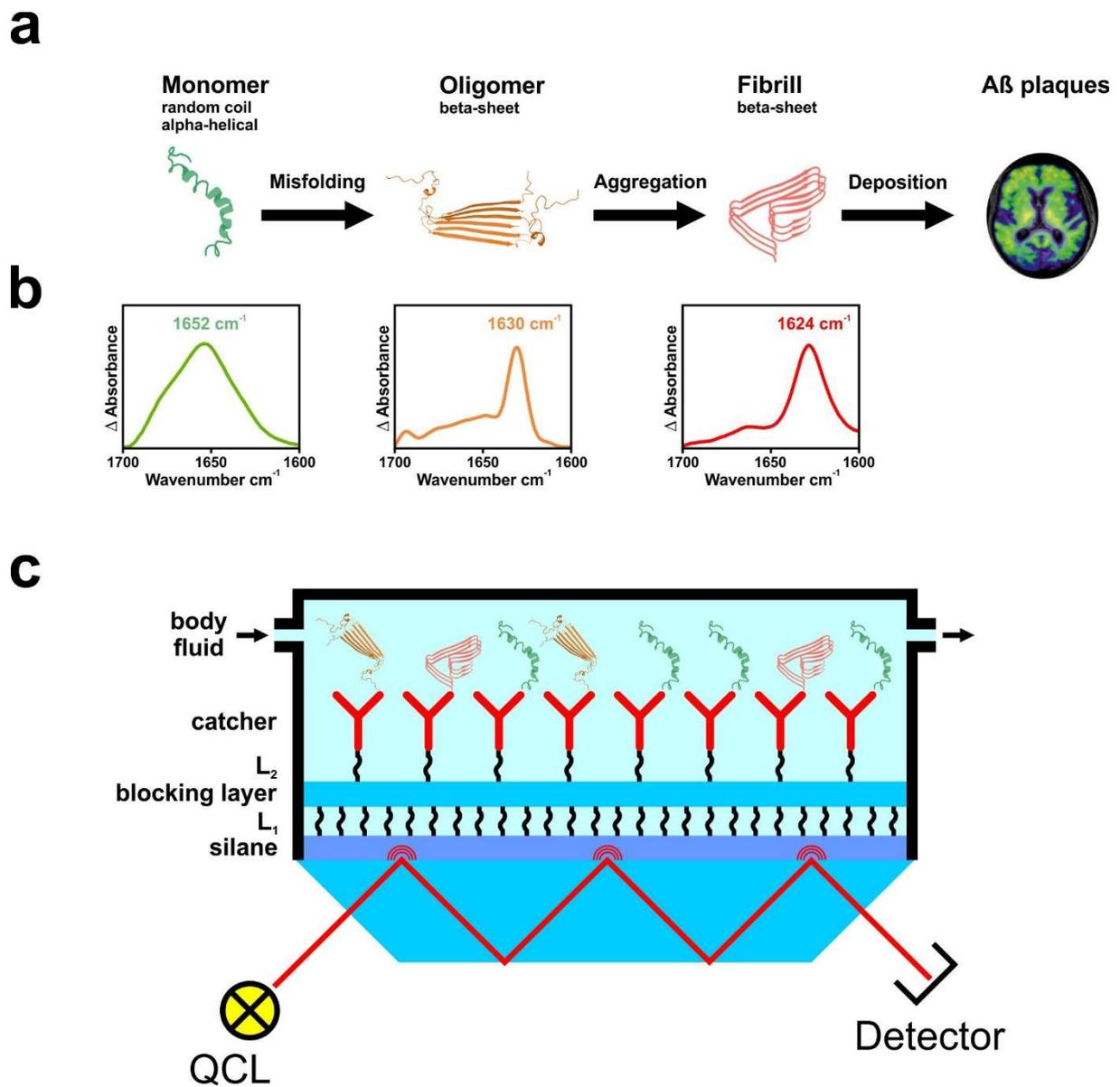

**Figure 1: The iRs concept: a** Progression of Alzheimer's disease: In the preclinical stage misfolding of Aβ leads first to oligomers and later to fibrils. Finally, Aβ is deposited in amyloid plaques which can be visualized by PET scanning. The Aβ plaques indicate the clinical stage of Alzheimer's disease. **b** Aβ Monomers (healthy), oligomers (at high risk) and fibrils (diseased) can be distinguished by their respective infrared absorbance maxima. The amide I maximum shifts from 1652 cm$^{-1}$ (monomers) to 1630 cm$^{-1}$ (oligomers) to 1624 cm$^{-1}$ (fibrils). **c** Functionalized surface with blocking layer, covalently attached by linker $L_1$ to silane covering the ATR-crystal surface and catcher antibodies covalently attached to the blocking layer via Linker $L_2$.

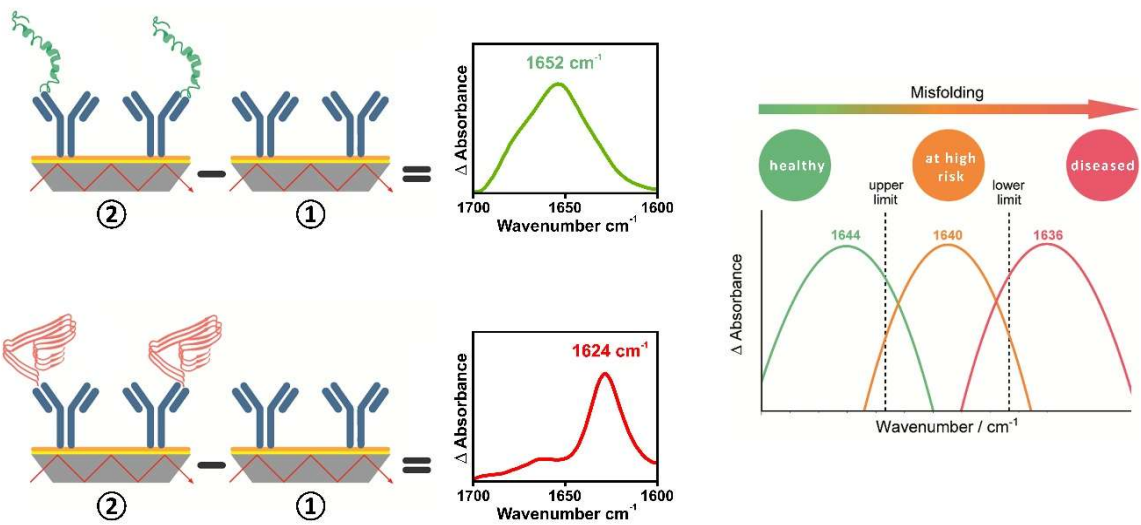

**Figure 2: Structural sensitive amide I band maximum:** The infrared-spectrum of the antigen is obtained by taking the aborbance-difference after (2) and before (1) the binding of the antigen. The amide I absorbance reflects the secondary structure composition of the antigen. The lower the maximum of the amide I band the higher the misfolding.

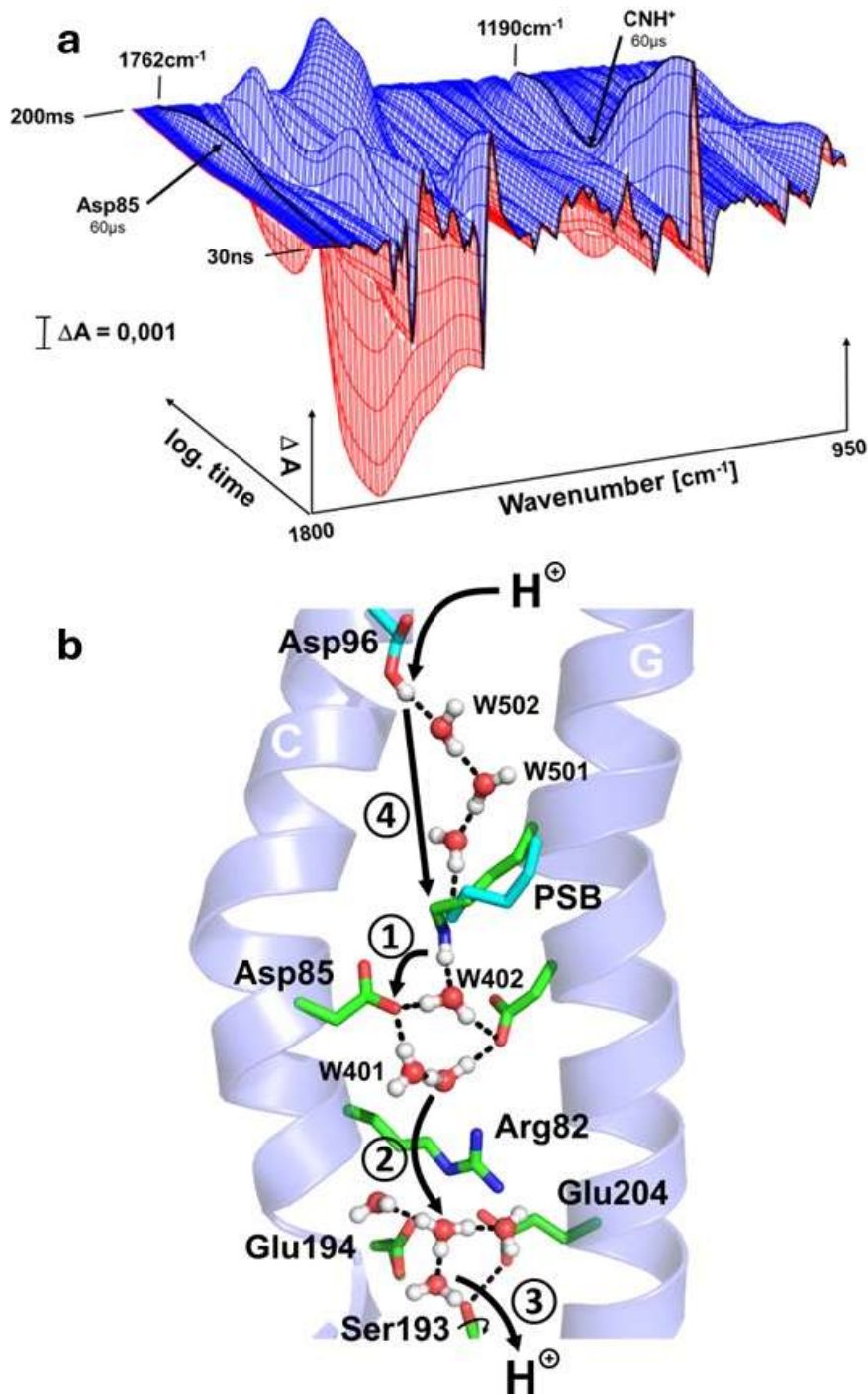

**Figure 3: Time-resolved FTIR spectroscopy:** Time-resolved FTIR-spectra taken during bacteriorhodopsins proton-pumping.[23–25] Difference-spectroscopy allows precise subtraction of a several orders of magnitude larger background absorbance to reveal time-resolved absorbance changes of the functional active protein-residues. Thereby, for example, the complete proton-transfer through a membrane protein via single water molecules and aspartic acids was resolved with a time-resolution from nanoseconds up to 200 milliseconds.

In **a** the time-resolved absorbance changes during proton pumping are shown. The decrease at 1190 cm$^{-1}$ indicates the Schiff base deprotonation and the increase at 1762 cm$^{-1}$ reflects protonation of the counterion Asp 85 in step 1 shown in **b**. Upon light activation, retinal undergoes isomerization from all-trans to 13-cis indicated at 1190 cm$^{-1}$. Subsequently, the proton from the Schiff base is transferred to Asp 85 in step 1. Arg 82 moves downwards in step 2 and induces release of an excess proton from a protonated Eigen-water cluster to the extracellular medium in step 3. Next (step 4), Asp96 reprotonates the Schiff base via a water chain formed transient for 1 ms through a Grotthuss-type mechanism. Finally, Asp96 is reprotonated from the cytoplasmic side and the retinal isomerizes back to all-trans.

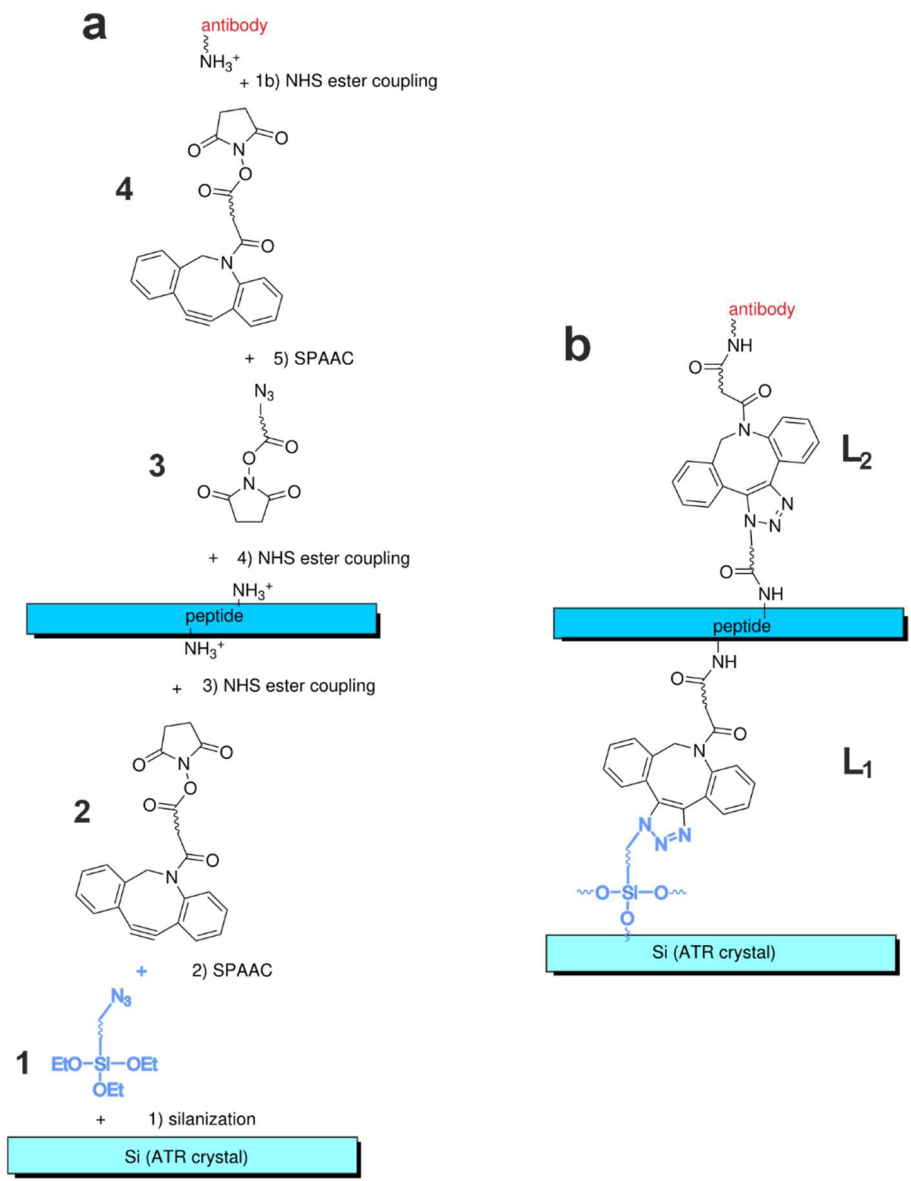

**Figure 4: Stepwise functionalization of the ATR surface: a** Initially, the ATR surface is coated with a triethoxysilane layer, followed by attachment of a blocking layer via the linker **2**. A second linker binds the capture antibody to this blocking layer. This results in the surface shown in b, where both the blocking layer and the antibody are covalently bound to the ATR crystal.

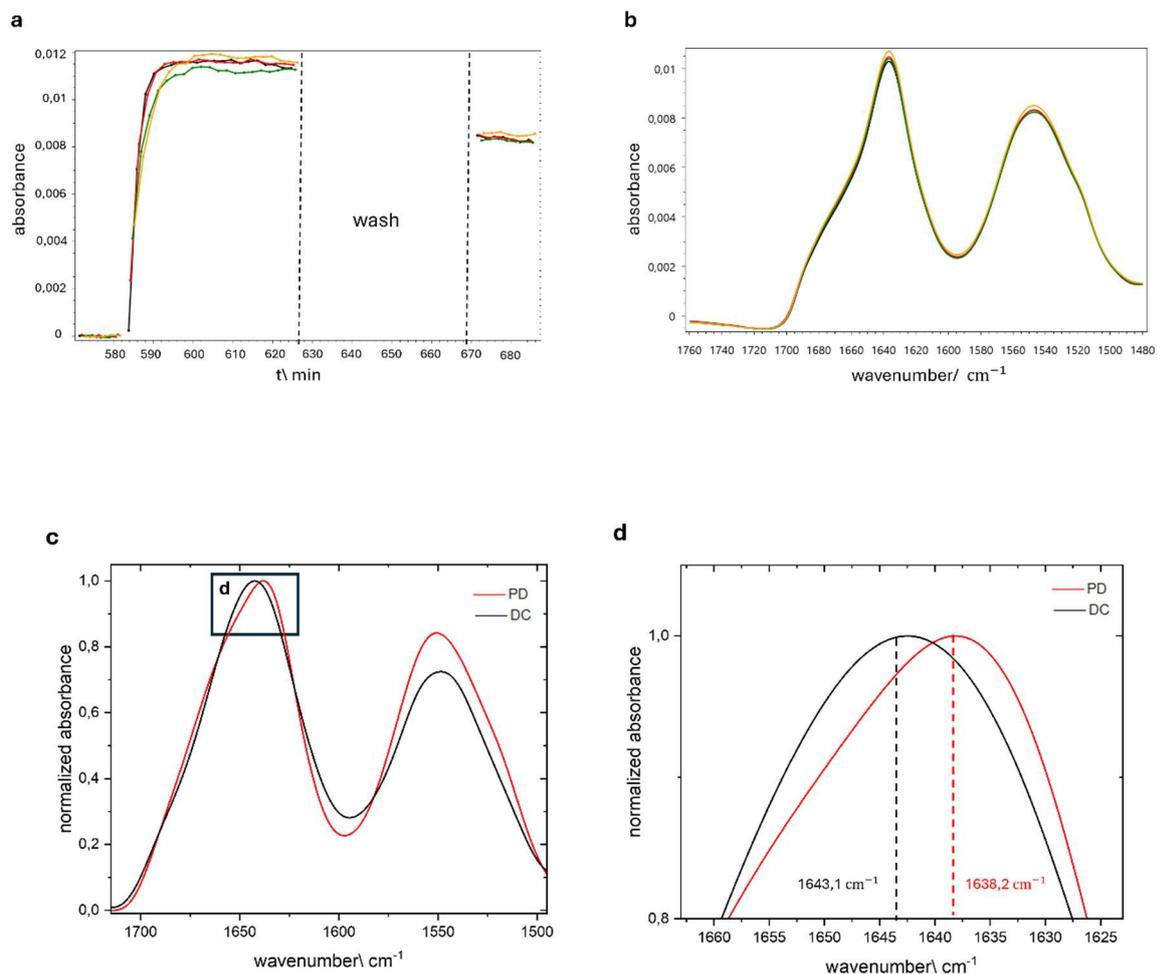

**Figure 5 Results of a typical iRS measurement: a** Kinetics of the amide II band during antibody immobilization in 4 different channels. After antibody binding a buffer washes off for about 40 minutes non-covalently absorbed antibody from the surface. The amide II band indicates the amount of antibody bound to the surface. **b** Amide I and II bands of the catcher antibody in 4 different channels on two different crystals. **c** Amide I and II band of a PD (Parkinson Disease) CSF sample and a DC (Disease Control) CSF sample measured with the QCL-based iRS. They reflect the binding of alpha-synuclein on the surface. All spectra were normalized by scaling the amide I band to an intensity value of 1. **d** Enlarged view of the amide I band maximum of the PD and DC spectra, respectively. The lower maximum for the PD sample indicates increased misfolding of alpha-synuclein compared to the DC sample. Study subjects gave their informed consent, and study approval was obtained by the local ethics committee and institutional review board (IRB Vote from Landesärztekammer Hessen: FF89/2008 and FF38/2016**)** in accordance with the Declaration of Helsinki**.**

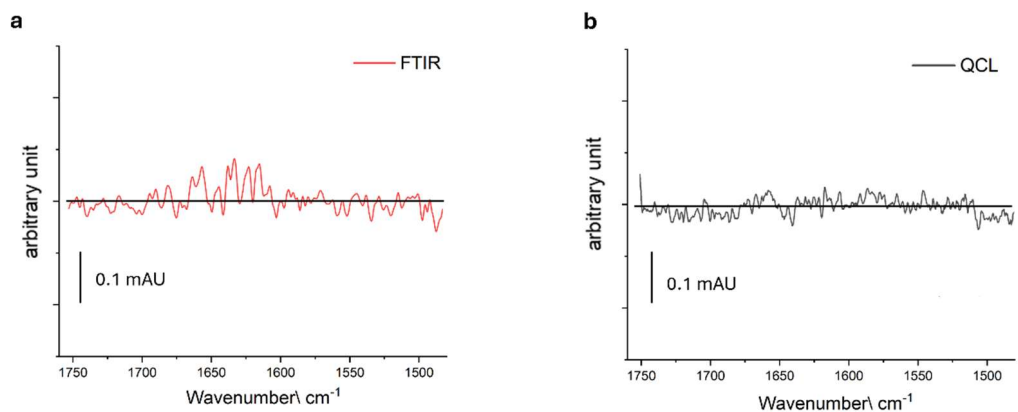

**Figure 6: S/N of FTIR and QCL based iRS instruments.** The FTIR data shown in **a** were obtained with a resolution of 2 cm⁻¹. For the QCL data shown in **b**, a moving average over 10 points was applied, also resulting in a 2 cm⁻¹ spectral resolution. Background and sample spectra were measured for 20 seconds each. The root mean square (RMS) noise was evaluated in the spectral region from 1610 to 1690 cm⁻¹, the amide I and OH bending absorbance region. The obtained RMS is $3.1 \cdot 10^{-5}$ for the FTIR (**a**) and $2.5 \cdot 10^{-5}$ for the QCL system (**b**).

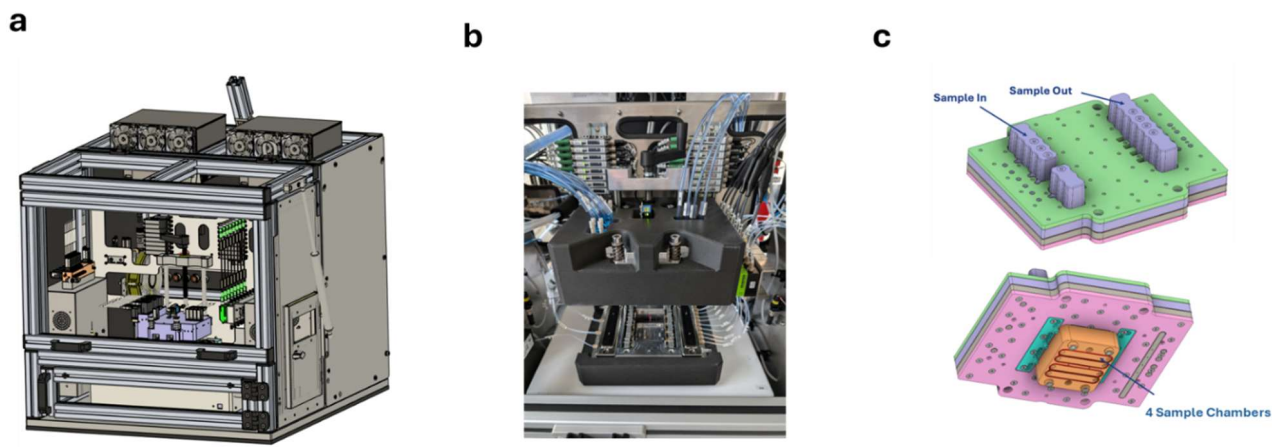

**Figure 7: QCL-based iRS System:** a fully automated CE-certified iRS instrument consisting of a QCL laser, IR-beam optics and detector in the lower compartment and a microfluidic system in the upper compartment. **b** microfluidic interface in an open state (the upper part is also seen in blue in **a**. The disposable is inserted between the upper and lower parts and connected to the advanced fluidic system. **c** Top view of the disposable with four input chambers for the body fluid samples. On the bottom view on the four-sample chamber (orange) two crystals each with two channels can be mounted to flow the body fluids over the functionalized crystal surfaces.